\documentclass[12pt]{article}

\usepackage{graphicx}
\usepackage{overcite}

\newcommand{\eref}[1]{(\ref{#1})}

\newcommand{\defn}{\textit}

\newlength{\figurewidth}
\renewcommand{\baselinestretch}{1.77}

\newcommand{\captionfonts}{\small}
\makeatletter
\long\def\@makecaption#1#2{%
  \vskip\abovecaptionskip
  \sbox\@tempboxa{{\captionfonts #1: #2}}%
  \ifdim \wd\@tempboxa >\hsize
    {\captionfonts #1: #2\par}
  \else
    \hbox to\hsize{\hfil\box\@tempboxa\hfil}%
  \fi
  \vskip\belowcaptionskip}
\makeatother

\begin{document}

{
\renewcommand{\baselinestretch}{1}

\title{A network analysis of committees in the\\
  United States House of Representatives}
\author{Mason A. Porter,$^{1,2}$\footnote{To whom correspondence should be 
addressed. E-mail: mason@math.gatech.edu} \, Peter J. Mucha,$^1$\, 
M. E. J. Newman,$^3$ and Casey M. Warmbrand$^4$\bigskip\\
  $^1$\footnotesize{School of Mathematics, Georgia Institute of
    Technology, Atlanta, GA
    30332--0160}\\
  $^2$\footnotesize{Center for Nonlinear Science, School of Physics,
    Georgia Institute of
    Technology, Atlanta, GA  30332--0430}\\
  $^3$\footnotesize{Department of Physics and Center for the Study of
    Complex Systems,}\\
  \footnotesize{University of Michigan, Ann Arbor, MI  48109--1120}\\
  $^4$\footnotesize{Department of Mathematics, University of Arizona,
    Tucson, AZ 85721--0089}}
    
\date{\today} 

\maketitle
}

\begin{abstract}
  Network theory provides a powerful tool for the representation and
  analysis of complex systems of interacting agents.  Here we
  investigate the United States House of Representatives network of
  committees and subcommittees, with committees connected according to
  ``interlocks'' or common membership.  Analysis of this network
  reveals clearly the strong links between different committees, as
  well as the intrinsic hierarchical structure within the House as a
  whole.  We show that network theory, combined with the analysis of
  roll call votes using singular value decomposition, successfully
  uncovers political and organizational correlations between
  committees in the House without the need to incorporate other
  political information.
\end{abstract}

\clearpage

\section*{Introduction}

Much of the detailed work in making United States law is performed by
Congressional committees and subcommittees.  This contrasts with
parliamentary democracies such as Great Britain and Canada, where a
larger part of the legislative process is directly in the hands of
political parties or is conducted in sessions of the entire
parliament.  While the legislation drafted by committees in the
U.S.~Congress is subject ultimately to roll call votes by the full
House and Senate, the important role played by committees and
subcommittees makes the study of their formation and composition vital
to understanding the work of the American legislature.

Several contrasting theories of committee assignment strategies have
been developed in the political science literature (mostly through
qualitative studies, although there have been some quantitative ones
as well) \cite{org,nisk,gill,kreh,cox,shep}, but there is no consensus
explanation of how committee assignments are initially determined or
of how they are modified from one session of Congress to the next.  A
question of particular interest is whether political parties assign
committee memberships essentially at random or if important
Congressional committees can be seen using objective analysis to be
``stacked'' with partisan party members.

The work presented here approaches these issues using a different set of
analytical tools from those previously employed.  We use the tools of
network theory, which have been successfully applied in recent years to
characterize a wide variety of complex systems.\cite{str01,newmansirev}.
As we show, network theory is particularly effective at uncovering
structure among committee and subcommittee assignments without the need to
incorporate any specific knowledge about committee members or their
political positions.

Although there has been only limited previous work on networks of
Congressional committees, there is a considerable body of literature
on other collaboration networks, such as the boards of directors of
corporations \cite{Mariolis75,Useem84,schwartz,robins,mizruchi}, which
occupy a position in the business world somewhat analogous to that
occupied by committees in Congress.  It has been shown that board
memberships and the networks they create play a major role in the
spread of attitudes, ideas, and practices through the corporate world,
affecting political donations \cite{Useem84}, investment strategies
\cite{Haunschild93}, and even the stock market on which a company is
listed \cite{RDW00}.  Studies of the structure of corporate networks
have shed considerable light on the mechanisms and pathways of
information diffusion \cite{ceo,burt,burt2}, and it seems plausible
that the structure of congressional committees would be similarly
revealing.

\section*{Networks of committees}

We study the U.S.~House of Representatives and construct bipartite or
``two-mode'' networks based on assignments of Representatives to
committees and subcommittees (henceforth called just ``committees''
for simplicity) in the 101st--108th Houses (1989--2004).
(Table~\ref{incharge} lists the House leadership during this period.)
These networks have two types of nodes, Representatives and
committees, with edges connecting each Representative to the
committees on which they sit.

\begin{figure}[t]
\begin{center}
\includegraphics[width = 15cm]{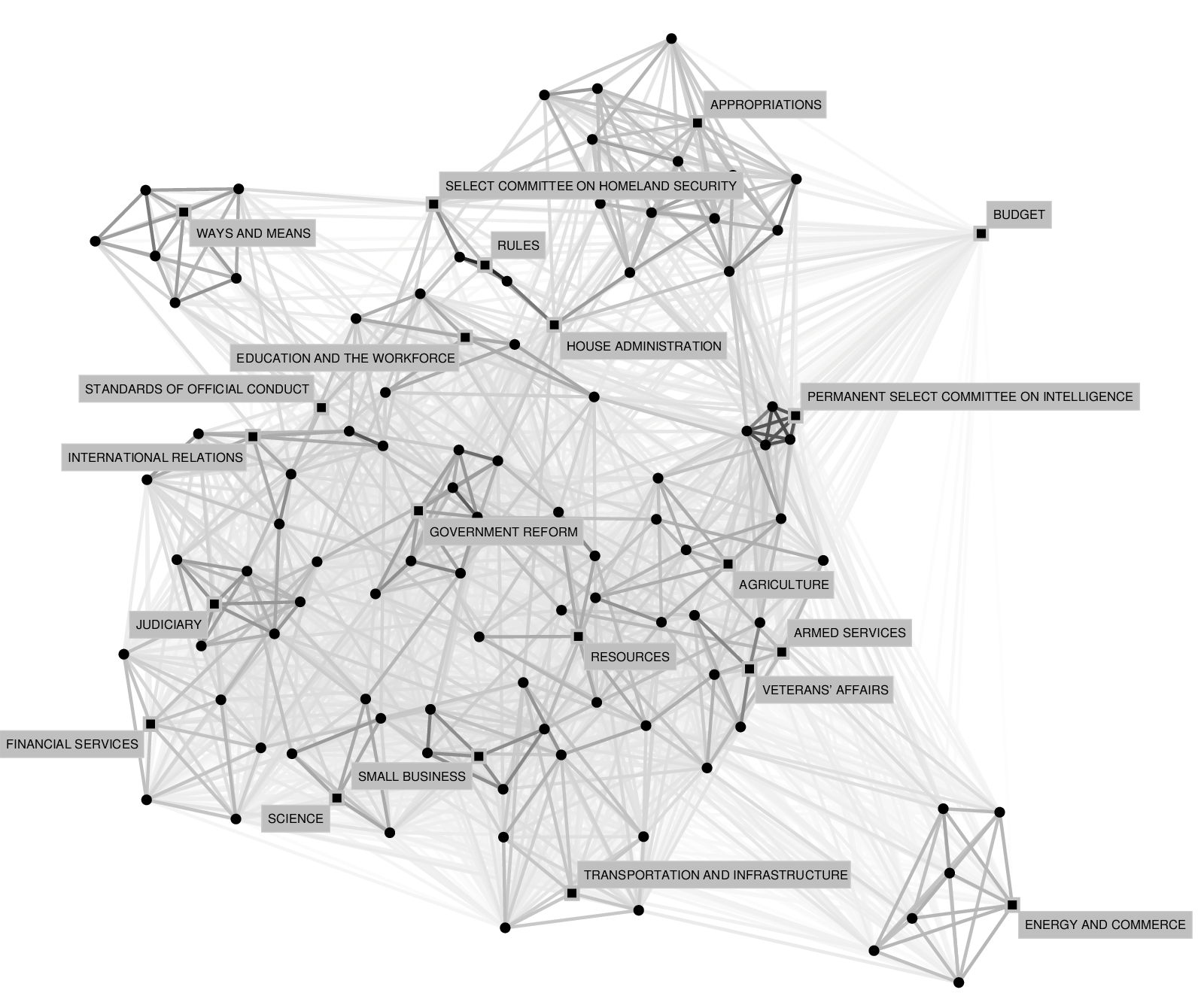}
\end{center}
\caption{Network of committees (squares) and subcommittees (circles) in the 
  107th U.S.~House of Representatives, with standing and select
  committees labeled.  (Subcommittees tend to be closely tied to their
  main committee and are therefore left unlabeled.)  Each link between
  two (sub)committees is assigned a strength equal to the normalized
  interlock.  Thus, lines between pairs of circles or pairs of squares
  represent normalized degree of joint membership between
  (sub)committees (it is because of this normalization that lines
  between squares are typically very light), and lines between squares
  and circles represent the fraction of standing committee members on
  subcommittees.  This figure is drawn using a variant of the
  Kamada-Kawai spring-embedding visualization, which takes link
  strengths into account.\cite{kk}}
\label{network}
\end{figure}

We project these two-mode committee assignment networks onto one-mode
networks whose nodes represent the committees and whose edges
represent common membership or ``interlocks'' between committees.
Figure~\ref{network} shows a visualization of the network of
committees for the 107th House of Representatives (2001--2002), an
example which we analyze in some depth.

The more common members two committees have, the stronger their
connection in the network.  We quantify the strength of connection by
the \defn{normalized interlock}, defined as the number of common members
divided by the expected number of such common members if committees of
the same size were randomly and independently assigned from the
members of the House.  Committees with as many common members as would
be expected by chance have normalized interlock~1, those with twice as
many have interlock~2, those with none have interlock~0, and so forth.


Some of the connections depicted in Fig.~\ref{network} are expected
and unsurprising.  For instance, one finds that sets of subcommittees
of the same larger committee share many of the same members, thereby
forming a group or clique in the network.  For example, the four
subcommittees of the 107th Permanent Select Committee on Intelligence
each include at least half of the full 20-member committee and at
least one third of each of the other subcommittees.  These tight
connections result in normalized interlocks with values in the range
14.4--23.6, which are substantially higher than average and cause
these five nodes to be drawn close together in the graph
visualization, forming the small pentagon in the middle right of
Fig.~\ref{network}.

We also find more surprising connections between committees.  For
instance, the 9-member Select Committee on Homeland Security, formed
in June 2002 during the 107th Congress in the aftermath of the
terrorist attacks of September 11, 2001,
is observed to have a strong connection to the 13-member Rules Committee
(with a normalized interlock of 7.4 from two common members), which is
the committee charged with deciding the rules and order of business
under which legislation will be considered by other committees and the
full House (see {\it thomas.loc.gov}). The Homeland Security Committee is also
connected to the 7-member Legislative and Budget Process Subcommittee
of Rules by the same two common members (with normalized interlock
13.7).  In the 108th Congress (not depicted here), the Homeland
Security Committee swelled to 50 members but maintained a close
association with the Rules Committee (with normalized interlock of 4.1
from 6 common members).

\section*{Structure of the House}

\begin{figure}[tp]
\begin{center}
  \includegraphics[width=15cm]{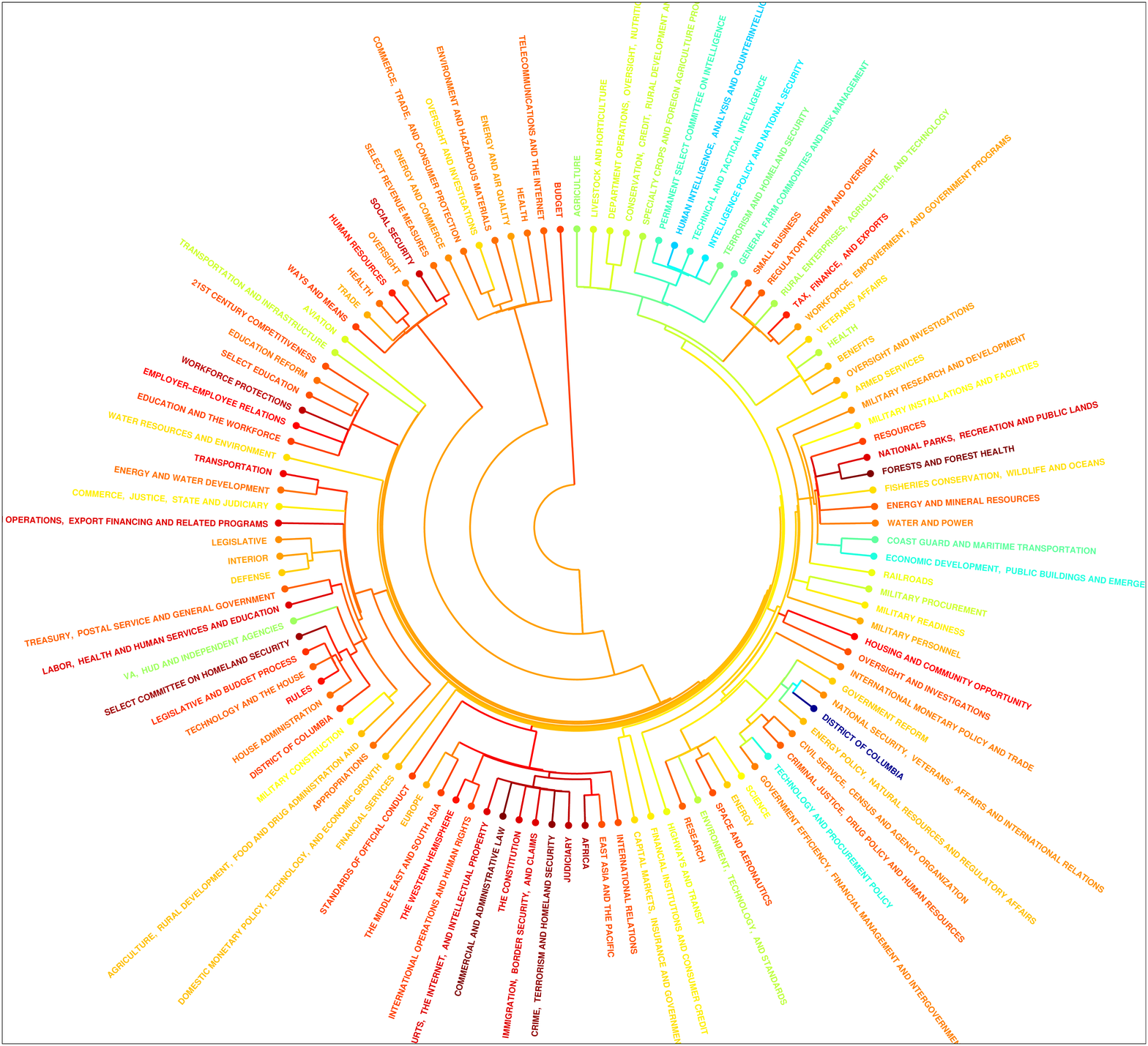}
\end{center}
\caption{Dendrogram representing the hierarchical clustering of the
committees of the 107th U.S.~House of Representatives, determined by
single-linkage clustering on normalized committee interlocks.  Each
committee is color-coded according to the mean ``extremism'' of its
members (defined in the main text), from less extreme (blue) to
more extreme (red).  The clusters at each level are color-coded
according to the average of their constituent committee extremism
scores.}
\label{107clust}
\end{figure}

Connections between committees can be quantified in greater detail by
applying the technique of \defn{single linkage
clustering}.\cite{cluster} Starting with the complete set of
committees for a given Congress, committees are joined together
sequentially starting with the pair with the greatest normalized
interlock, followed by the next greatest, and so forth.  This process
generates ``clusters'' of committees, which can be represented using a
tree or \defn{dendrogram}, such as that shown in Fig.~\ref{107clust}
for the 107th House.  There appear to be essentially four hierarchical
levels of clusters within this dendrogram: subcommittees, committees,
groups of committees, and finally the entire House.\cite{strahler}
(There is also some indication of a weaker fifth level of organization
corresponding to groups of subcommittees inside larger standing
committees.)

Here, we are primarily interested in the third hierarchical
level---the connections between committees.  For example, we see near
the 8 o'clock position in Fig.~\ref{107clust} a tightly grouped
cluster that includes the House Rules Committee and the Select
Committee on Homeland Security.  Because assignments to select
committees are ordinarily determined by drawing \emph{selectively}
from legislative bodies with overlapping jurisdiction, one might naively 
expect a close connection between the Select
Committee on Homeland Security and, for example, the Terrorism and
Homeland Security Subcommittee of the Intelligence Select Committee,
which was formed originally as a bipartisan ``working group'' and was
designated on September 13, 2001 by House Speaker Dennis Hastert
[R-IL] as the lead congressional entity assigned to investigate the
9/11 terrorist attacks (see 
{\it www.homelandsecurity.org/journal/articles/kaniewskilegislative.htm}).  
However, the 107th Homeland
Security Committee shares only one common member (normalized interlock
2.4) with the Intelligence Select Committee (located near the 1
o'clock position in Fig.~\ref{107clust}) and has no interlock with any
of the four Intelligence subcommittees.

\section*{Voting patterns}

A further twist can be introduced by considering how the network of
interlocks between committees is related to the political positions of
their constituent Representatives.  One way to characterize political
positions is to tabulate individuals' voting records on key issues,
but such a method is subjective by nature and a method that involves
less personal judgment on the observer's part is preferable.  Here we
use a singular value decomposition (SVD)\cite{vanloan} of voting
records.\cite{pr,pr00,sirovich}  Other data mining methods can also be used 
(see {\it http://www.ailab.si/aleks/Politics/}).

We define an $n\times m$ \defn{voting matrix} $\mathbf{B}$ with one
row for each of the $n$ Representatives in the House and one column
for each of the $m$ votes taken during the session.  For instance, the
107th House had $n=444$ Representatives (including mid-term
replacements) and took $m=990$ roll call votes.  The element $B_{ij}$
is $+1$ if Representative~$i$ voted ``yea'' on vote~$j$ and $-1$ if he
or she voted ``nay''.  If a Representative did not vote because of
absence or abstention, the corresponding element is~$0$.

The SVD identifies groups of Representatives who voted in a similar
fashion on many votes.  The grouping that has the largest mean-square
overlap with the actual groups voting for or against each measure is
given by the leading (normalized) eigenvector $\mathbf{u}^{(1)}$ of
the matrix~$\mathbf{B}^T\mathbf{B}$, the next largest by the second
eigenvector, and so on \cite{sirovich,vanloan}.  If we denote by
$\sigma_k^2$ the corresponding eigenvalues (which are provably
non-negative) and by $\mathbf{v}^{(k)}$ the normalized eigenvectors of
$\mathbf{B}\mathbf{B}^T$ (which have the same eigenvalues), then it
can be shown that
\begin{equation}
B_{ij} = \sum_{k=1}^n \sigma_k u^{(k)}_i v^{(k)}_j,
\label{decomposition}
\end{equation}
and that the matrix~$\mathbf{B}^{(r)}$ with elements
\begin{equation}
B^{(r)}_{ij} = \sum_{k=1}^r \sigma_k u^{(k)}_i v^{(k)}_j,
\label{full}
\end{equation}
approximates the full voting matrix~$\mathbf{B}$, with the sum of the
squares of the errors on the elements equal to $\sum_{k=r+1}^n
\sigma_k^2$, which vanishes in the limit $r\to n$.  Assuming the
quantities $\sigma_k$, which are called the \defn{singular values},
are ordered such that $\sigma_1\ge\sigma_2\ge\sigma_3\ldots$, this
means that $\mathbf{B}^{(r)}$ will be an excellent approximation to
the original voting matrix if the singular values fall off
sufficiently rapidly with increasing~$k$.

\begin{figure}[t]
\begin{center}
  \includegraphics[width=8.7cm]{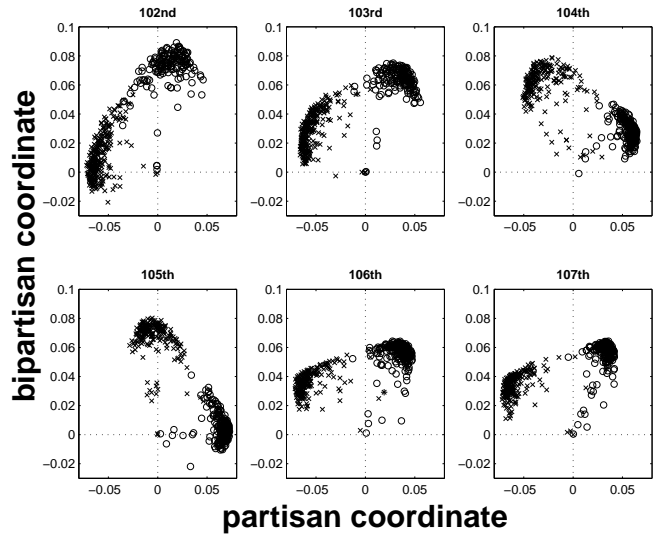}
\end{center}
\caption{Singular value decomposition (SVD) of the voting record of the 
  House of Representatives in the 102nd--107th U.S.~Congresses.  Each
  point represents a projection of a Representative's votes onto
  eigenvectors corresponding to the leading two singular values.  The
  two axes are denoted ``partisan'' and ``bipartisan,'' as described
  in the text.  Democrats (x) appear on the left, whereas Republicans
  (o) are on the right. The few independents are marked by asterisks
  (*).}
\label{svdeg}
\end{figure}

Alternatively, one can say that the $k$th term in the singular value
decomposition \eref{decomposition} accounts for a fraction
$\sigma_k^2/\sum_{k=1}^n \sigma_k^2$ of the sum of the squares of the
elements of the voting matrix.  For the 107th House, we find that the
leading eigenvector accounts for about 45.3\% of the voting matrix, the
second eigenvector accounts for about 29.6\%, and no other eigenvector
accounts for more than 1.6\%.  Thus, to an excellent approximation, a
Representative's voting record can be characterized by just two
coordinates, measuring the extent to which they align (or do not align)
with the groups represented by the first two eigenvectors.  That is,
\begin{equation}
B^{(2)}_{ij} = \sigma_1 u^{(1)}_i v^{(1)}_j + \sigma_2 u^{(2)}_i v^{(2)}_j
\end{equation}
is a good approximation to $B_{ij}$.  Similar results are obtained for
other sessions of Congress, with two eigenvectors giving a good
approximation to the voting matrix in every case.  It has been shown
previously using other methods that Congressional voting positions are
well-approximated by just two coordinates (see {\it voteview.com}), but the SVD
does so in a particularly simple fashion directly from the roll call data.
In Fig.~\ref{svdeg}, we
plot the two coordinates for every member of the House of
Representatives for each of the 102nd--107th Congresses.

We find that the leading eigenvector\footnote{This holds for the
  101st--105th Houses.  The leading and second eigenvectors switch
  roles in the 106th and 107th Houses.} corresponds closely to the
acknowledged political party affiliation of the Representatives, with
Democrats ($\times$) on the left and Republicans ($\circ$) on the
right in the plots.  We therefore call this the ``partisan
coordinate'' and Representatives who score highly on it---either
positively or negatively---tend often to vote with members of their
own party.  From the partisan coordinate, we also compute a measure of
``extremism'' for each Representative as the absolute value of their
partisan coordinate relative to the mean partisan score of the full
House.  That is, we define the extremism $e_i$ of a Representative by
$e_i = |p_i - \mu|$, where $p_i$ is the Representative's partisan
coordinate and $\mu$ is the mean of that coordinate for the entire
House (which is usually skewed slightly towards the majority party).
In Table \ref{part}, we list the most and least partisan
Representatives from each party computed from the roll call of the
107th House.  We also compare our rank orderings to those obtained
using an alternative method, the Optimal Classification (OC) technique
of Poole and Rosenthal\cite{pr00} (also applied only to the
107th House).

In contrast, the second eigenvector groups essentially all
Representatives together regardless of party affiliation and thus
appears to represent voting actions in which most members of the House
either approve or disapprove of a motion simultaneously.  We call this
the ``bipartisan coordinate,'' as Representatives who score highly on
it tend often to vote with the majority of the House.

\begin{figure}[t]
\begin{center}
  \includegraphics[width=8.7cm]{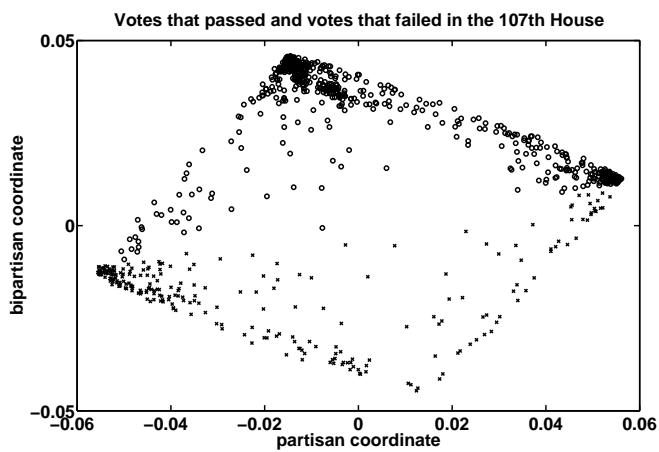}
\end{center}
\caption{SVD of the roll call of the 107th House projected onto the voting 
  coordinates.  Points represent projections of the votes cast on a
  measure onto eigenvectors corresponding to the leading two singular
  values. There is a clear separation between measures that passed
  (o) and those that did not (x).  The four corners of the plot
  are interpreted as follows: measures with broad bipartisan support
  (north) all passed; those supported mostly by the Right (east)
  passed because the Republicans constituted the majority party of the
  107th House; measures supported by the Left (west) failed because of
  the Democratic minority; and the (obviously) very few measures
  supported by almost nobody (south) also failed.}
\label{votesvd}
\end{figure}

Using our SVD results, we can also calculate the positions of the
\emph{votes} (as opposed to the voters) along the same two leading
dimensions to quantify the nature of the issues being decided.  We
show this for the 107th House in Fig.~\ref{votesvd}.  One application
of this analysis is a measurement of the reproducibility of individual
votes and outcomes.  By reconstituting the voting matrix using only
the information contained in the two leading singular values and the
corresponding eigenvectors and subsequently summing the resulting
approximated votes over all Representatives, we derive a single score
for each vote.  Making a simple assignment of ``pass'' to those votes
that have a positive score and ``fail'' to all others
successfully reconstructs the outcome of $984$ of the $990$ total
votes, or about 99.4\%.  (Overall, $735$, or about 74.2\%, of the
votes passed, so simply guessing that every vote passed would be
considerably less effective.)  If we throw out known absences and
abstentions, the analysis still identifies $975$ of the $990$ results
correctly.  Even with the most conservative measure of this
computation's success rate, in which we throw out abstentions and
absences first and then examine \emph{individual} Representatives'
yeas/nays (approximately 92.7\% of which are correctly identified by
the sign of the elements in the projection of the voting matrix), the
two-dimensional reconstruction still identifies $939$, or about
94.8\%, of the votes correctly.  We have repeated these calculations
for the 101st--106th Houses and found similar results in each case.
(The remarkable accuracy of SVDs in reconstructing votes was
previously observed for the example of the U.S.~Supreme Court \cite{sirovich}.)

The SVD analysis gives a simple way of classifying Representatives'
voting positions without making subjective judgments.  In
Fig.~\ref{107clust}, we have combined our clustering analyses of
committees with the SVD results by color-coding each committee
according to the mean ``extremism'' of its members, so that committees
populated by highly partisan members of either party appear in red and
committees containing more moderate Representatives appear in blue.
Taking again the examples of Intelligence and Homeland Security, the
figures immediately identify the former as moderate and the latter as
more partisan.  Indeed, the Select Committee on Homeland Security has
a larger mean extremism than any of the 19 standing Committees and has
the 4th largest mean extremism among the 113 committees of the 107th
House.  This is perhaps not so surprising when we see that its
constituent Representatives included the House Majority Leader,
Richard Armey [R-TX], and both the Majority and Minority Whips, Tom
DeLay [R-TX] and Nancy Pelosi [D-CA].  However, this characterization
of the committee was made \emph{mathematically}, using no knowledge
beyond the roll call votes of the 107th House.  As another example,
the 107th House Rules Committee is the 2nd most extreme of the 19
standing committees (after Judiciary) and ranks 18th out of 113
committees overall.  By contrast, the Permanent Select Committee on
Intelligence of the 107th House has a smaller mean extremism than any
of the 19 standing Committees, and Intelligence and its four
subcommittees all rank among the 10 least extreme of all 113
committees.

\section*{Conclusions}

To conclude, a network theory approach coupled with an SVD analysis of
roll call votes is demonstrably useful in analyzing organizational
structure in the committees of the U.S.~House of Representatives.  We
have found evidence of several levels of hierarchy within the network
of committees and---without incorporating any knowledge of political
events or positions---identified some close connections between
committees, such as that between the House Rules Committee and the
Select Committee on Homeland Security in the 107th Congress, as well
as correlations between committee assignments and Representatives'
political positions.  Our analysis of committee interlocks and voting
patterns strongly suggests that committee assignments are not
determined at random (i.e., that some committees are indeed
``stacked'') and also indicates the degree of departure from
randomness.  We have discussed here a few observations in detail, but
a rich variety of other results can be derived from similar analyses.
We hope that further studies using similar techniques will provide key
insights into the structure of the House of Representatives and other
political bodies.

\section*{Acknowledgements}

We gratefully acknowledge Thomas Callaghan, Michael Cohen, Gordon
Kingsley, and Scott Page for useful conversations during this
research.  This work was supported in part by the National Science
Foundation, including CMW's contribution through an NSF VIGRE Research
Experiences for Undergraduates program and MAP's visiting faculty
position at Georgia Tech.  We also thank Michael Abraham for
developing some of the computer code used in this research and Thomas
Bartsch, Steven Lansel, Tom Maccarone, Slaven Pele\v{s}, Steve Van
Hooser, and Julia Wang for critical readings of the manuscript.  We
also thank the two anonymous referees whose constructive comments
improved this manuscript greatly.  We obtained the roll calls for the
102nd--107th Congresses from the Voteview website ({\it voteview.com}) the
roll calls for the 101st Congress from the Inter-University Consortium
for Political and Social Research (see {\it www.icpsr.umich.edu}) and the committee
assignments for the 101st--108th Congresses from the web site of the
House of Representatives Office of the Clerk (see {\it clerk.house.gov}).

\bibliographystyle{pnas}

\clearpage{}

\clearpage{}

\begin{table}[t] 
\centerline{
{\footnotesize
\begin{tabular}{|c|c|c|c|c|c|} \hline
Congress & Speaker & Majority Leader & Minority Leader & Majority Whip & Minority Whip \\ \hline
101st (89--90) & T.~S. Foley & R.~A. Gephardt & R.~H. Michel & T. Coelho, W.~H. Gray III &  D. Cheney, N.~L. Gingrich \\
102nd (91--92) & T.~S. Foley & R.~A. Gephardt & R.~H. Michel & W.~H. Gray III, D.~E. Bonior & N.~L. Gingrich \\
103rd (93--94) & T.~S. Foley & R.~A. Gephardt & R.~H. Michel & D.~E. Bonior & N.~L. Gingrich \\
104th (95--96) & N.~L. Gingrich & R.~K. Armey & R.~A. Gephardt & T.~D. DeLay & D.~E. Bonior \\
105th (97--98) & N.~L. Gingrich & R.~K. Armey & R.~A. Gephardt & T.~D. DeLay & D.~E. Bonior \\
106th (99--00) & J.~D. Hastert & R.~K. Armey & R.~A. Gephardt & T.~D. DeLay & D.~E. Bonior \\ 
107th (01--02) & J.~D. Hastert & R.~K. Armey & R.~A. Gephardt & T.~D. DeLay & N. Pelosi \\
108th (03--04) & J.~D. Hastert & T.~D. DeLay & N. Pelosi & R. Blunt & S. Hoyer \\ \hline
\end{tabular}}}
\caption{United States House of Representatives leadership from the 
101st--108th Congresses.  The Democrats held the House majority in the 
101st--103rd Congresses (1989--1994), and the Republicans held it in the 
104th--108th Congresses (1995--2004).}
\label{incharge}
\end{table}

\begin{table}[t] 
\centerline{
{\footnotesize
\begin{tabular}{|c|c|c|} \hline
Least Partisan & Farthest Left & Farthest Right \\ \hline
K. Lucas [R] & J.~D. Schakowsky [D] & T.~G. Tancredo [R] \\
C.~A. Morella [R] & J.~P. McGovern [D]  & J.~B. Shadegg [R]\\
R.~M. Hall [D] & H.~L. Solis [D] & J. Ryun [R] \\
R. Shows [D] & L.~C. Woolsey [D] & B. Schaffer [R] \\
G. Taylor [R] & J.~F. Tierney [D] & P. Sessions [R]  \\
C.~W. Stenholm [D] & S. Farr [D] & S. Johnson [R]  \\ 
R.~E. Cramer [D]  & N. Pelosi [D]   & B.~D. Kerns [R]  \\
V.~H. Goode [R] & E.~J. Markey [D]  & P.~M. Crane [R] \\ 
C. John [D] & J.~W. Olver [D] & W.~T. Akin [R] \\
C.~C. Peterson [D] & L. Roybal-Allard [D] & J.~D. Hayworth [R] \\ \hline
\end{tabular}}}
\caption{SVD rank ordering of the most and least partisan Representatives in 
the 107th U.S.~House.  The 1st column gives the least partisan Representatives,
 as determined by an SVD of the roll call votes.  The 2nd column gives the SVD 
rank ordering of the most partisan Representatives.  They are all Democrats, so
 this also gives the rank of the Representatives farthest to the Left.  The 3rd
 column gives the rank of the Representatives farthest to the Right.  The SVD 
rank ordering was determined for Representatives after mid-term replacements 
(432 total Congressmen) using all 990 roll calls; it classifies 92.7\% of 
individual votes correctly.  By contrast, in Poole and Rosenthal's Optimal 
Classification (OC) method \cite{pr00}, a rank ordering of the 
Representatives in the 107th House is determined using 443 total 
Representatives and 749/990 roll calls (votes with fewer than 0.5\% of the 
votes in the minority were removed from consideration).  It classifies 92.8\% 
of the individual Representatives' votes correctly.}  



\label{part}
\end{table}

\end{document}